\documentclass[fleqn,usenatbib]{mnras}

\usepackage{newtxtext,newtxmath}

\usepackage[T1]{fontenc}

\DeclareRobustCommand{\VAN}[3]{#2}
\let\VANthebibliography\thebibliography
\def\thebibliography{\DeclareRobustCommand{\VAN}[3]{##3}\VANthebibliography}

\usepackage{graphicx}	
\usepackage{amsmath}	

\usepackage{bm}
\usepackage{array}

\newcommand{\be}{\begin{equation}}
\newcommand{\ee}{\end{equation}}
\newcommand{\bea}{\begin{eqnarray}}
\newcommand{\eea}{\end{eqnarray}}
\newcommand{\dd}{\mathrm{d}}

\title[Molecular cloud mergers in the galactic context]{Building the molecular cloud population: the role of cloud mergers}

\author[Skarbinski, Jeffreson and Goodman]{
Maya~Skarbinski$^1$\thanks{mayaskarbinski@college.harvard.edu}, Sarah~M.~R.~Jeffreson$^1$\thanks{sarah.jeffreson@cfa.harvard.edu} and Alyssa A.~Goodman$^{1,2}$
\\
$^{1}$ Center for Astrophysics, Harvard \& Smithsonian, 60 Garden St, Cambridge, MA 02138, USA \\
$^{2}$ Radcliffe Institute for Advanced Study, Harvard University, 10 Garden St, Cambridge, MA 02138, USA \\
}

\date{Accepted 2022 November 25. Received 2022 November 03; in original form 2022 August 25}

\pubyear{2021}

\begin{document}
\label{firstpage}
\pagerange{\pageref{firstpage}--\pageref{lastpage}}
\maketitle

\begin{abstract}
We study the physical drivers of slow molecular cloud mergers within a simulation of a Milky Way-like galaxy in the moving-mesh code {\sc Arepo}, and determine the influence of these mergers on the mass distribution and star formation efficiency of the galactic cloud population. We find that $83$~per~cent of these mergers occur at a relative velocity below $5$~km/s, and are associated with large-scale atomic gas flows, driven primarily by (1) expanding bubbles of hot, ionised gas caused by supernova explosions and (2) galactic rotation. The major effect of these mergers is to aggregate molecular mass into higher-mass clouds: mergers account for over 50~per~cent of the molecular mass contained in clouds of mass $M>2 \times 10^6~{\rm M}_\odot$. These high-mass clouds have higher densities, internal velocity dispersions and instantaneous star formation efficiencies than their unmerged, lower-mass precursors. As such, the mean instantaneous star formation efficiency in our simulated galaxy, with its merger rate of just 1~per~cent of clouds per Myr, is $25$~per~cent higher than in a similar population of clouds containing no mergers.
\end{abstract}



\section{Introduction} \label{Sec::Introduction}
As the sites of galactic star formation, the evolution of giant molecular clouds is deeply intertwined with the rate and efficiency of star formation in galaxies. Molecular clouds are rapidly-evolving, with lifetimes much shorter than the orbital period of the host galaxy~\citep{Engargiola03,Blitz2007,Kawamura09,Murray11,2012ApJ...761...37M,Chevance20}. Cloud properties such as the internal velocity dispersion and surface density~\citep[e.g.][]{Leroy17,Sun18,Sun2020}, the dense gas fraction~\citep{Usero15,Bigiel16} and the star formation efficiency per free-fall time~\citep{Utomo18}, are observed to vary with the large-scale galactic environment. These environmental variations have been tied to the response of molecular gas to large-scale galactic-dynamical processes, and to the disruption and ionisation of molecular gas by stellar feedback~\citep[e.g.][]{Semenov17,Meidt18,2020MNRAS.498..385J,2021MNRAS.505.4048L}.

Molecular cloud mergers are one dynamically-driven process that may significantly alter the physical properties of the galactic cloud population and its star formation. Their influence is most often studied in the context of triggered star formation~\citep[e.g.][among many others]{Tan00,TaskerTan09,2012MNRAS.424..377D,2014MNRAS.439..936F,2022MNRAS.tmp.1143L}. That is, it is assumed that a significant fraction of merging clouds collide at velocities substantially greater than their internal velocity dispersions, such that a shockwave is formed, compressing the gas within the merged cloud to high densities, and leaving a burst of star and cluster formation in its wake. Simulations of individual cloud mergers, reaching mass resolutions below one solar mass, confirm that star formation may be triggered in these `fast' mergers, with ratios of collision velocity to internal cloud velocity dispersion that are greater than $\sim 3$~\citep[e.g.][]{2015MNRAS.453.2471B,Balfour17,2020MNRAS.499.1099L,2021arXiv210906195H}. However, it is debated whether (1) the bursts of triggered star formation are large-enough and of long-enough duration to significantly enhance the per-cloud star formation rate over its lifetime~\citep{2021arXiv210906195H}, and (2) whether enough high-speed mergers occur to substantially affect the galactic cloud population~\citep{Jeffreson+Kruijssen18}.

By contrast, the influence of `slow' cloud mergers, with collision speeds comparable to, or lower than, the internal velocity dispersion of the merging clouds, have not been studied as extensively. Simulations of flocculent spiral galaxies~\citep[e.g.][]{2015MNRAS.446.3608D,2021MNRAS.505.1678J} indicate that such `slow' mergers may account for the majority of cloud mergers in galaxies without bars or grand-design spiral patterns, and hypothesise that their primary influence is to aggregate mass into higher-mass clouds. The analytic theory of~\cite{2017ApJ...836..175K}, which predicts the form of galactic molecular cloud mass functions (assuming fixed time-scales of $\sim 10$~Myr for cloud accretion, star formation and dispersal) shows that the occurrence of cloud mergers may substantially increase the number of the highest-mass molecular clouds. Both observations~\citep{Murray11} and simulations~\citep{2011ApJ...738..101G,Jeffreson22} demonstrate that the highest-mass molecular clouds have the highest star formation efficiencies and account for the majority of galactic-scale star formation, and so the presence of cloud mergers at any collision speed may systematically alter the galactic-scale star formation efficiency, independently of triggered star formation.

In contrast to the `fast' cloud mergers at speeds $>10$~km/s, which must be driven by large-scale galactic rotation or shearing within galactic bars and spiral arms, `slow' mergers may also be produced by the converging gas flows associated with feedback-driven bubbles in the interstellar medium. Recently, three-dimensional spatial and kinematic data from our Solar neighbourhood have revealed molecular clouds arranged on the surfaces of such giant supernova-driven bubbles~\citep{2021ApJ...919L...5B,2022Natur.601..334Z,Foley22}: a constraint of $7$~km/s on the expansion velocity of the molecular gas on the surface of the the Local Bubble is given by~\cite{2022Natur.601..334Z}. Such large, feedback-driven bubbles can also be seen in simulations of isolated galaxies at spatial resolutions of a few parsecs~\cite[e.g.][]{Dobbs15,Tress20,2021MNRAS.505.3470J}, and in cosmological zoom-in simulations~\citep[e.g.][]{2020MNRAS.497.3993B}. The occurrence of cloud mergers on the surfaces of such bubbles is accounted for in many existing theoretical works~\citep[e.g.][]{2015A&A...580A..49I,2017ApJ...836..175K}, including the standard picture of the three-phase, supernova-driven, interstellar medium~\citep{1977ApJ...218..148M}.

In this paper, we investigate the driving of molecular cloud mergers, and their influence on cloud properties, across the disc of an entire flocculent spiral galaxy. Because the majority of mergers in such flocculent galaxies are `slow' mergers, we can study their influence on galactic-scale star formation without requiring that triggered star formation be resolved. We seek to answer the following three questions: \textbf{(1) what physical processes drive these mergers, and at what rate? (2) is the primary effect of these `slow' mergers to aggregate mass into high-mass molecular clouds? and (3) what effect does this mass aggregation have on the galactic cloud mass function, and on the galactic star formation efficiency?} After introducing the simulation in Section~\ref{Sec::sims}, we answer these three questions in Sections~\ref{Sec::driving}, \ref{Sec::before-and-after} and \ref{Sec::merger-influence}, respectively. We discuss our results and their caveats in the context of the literature in Section~\ref{Sec::discussion}, and summarise our conclusions in Section~\ref{Sec::conclusions}.

\section{Numerical methods} \label{Sec::sims}
In this section we provide the details of our numerical simulation of a Milky Way-like galaxy, our technique for molecular cloud identification and tracking, and our technique for identifying the sample of molecular cloud mergers analysed in this work.

\subsection{Simulation} \label{Sec::simulation}
\begin{figure}
	\includegraphics[width=\columnwidth]{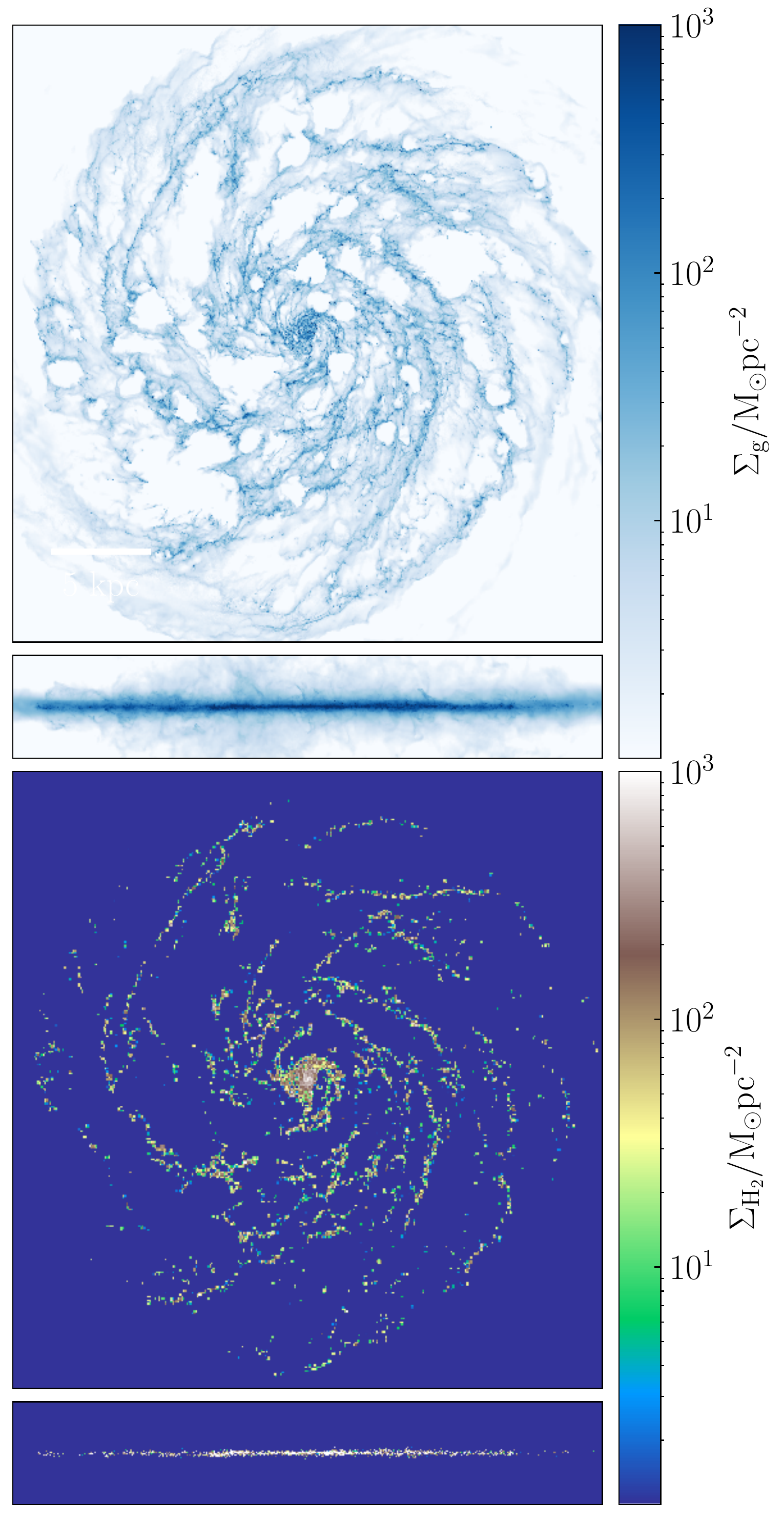}
	\caption{Column density maps of the total (upper two panels) and molecular (lower two panels) gas distribution for the simulated galaxy, at a simulation time of $600$~Myr.}
	\label{Fig::morphology}
\end{figure}

The isolated disc galaxy simulation presented in this work was first described in~\cite{2021MNRAS.505.3470J} (and named `HII heat and beamed mom' in that work). The spatial distribution of the total (upper panels) and molecular (lower panels) gas reservoirs is shown at the face-on and edge-on viewing angles in Figure~\ref{Fig::morphology}. Here we give an overview of the key numerical parameters, and refer the reader to the cited work for a fuller explanation.

Our simulation begins with the initial condition generated for the Agora comparison project~\citep{Kim14}, which resembles the Milky Way at redshift $z \sim 0$. The dark matter halo follows the form of~\cite{Navarro97}, with mass $M_{200} = 1.07 \times 10^{12}~{\rm M}_\odot$, virial radius $R_{200} = 205~{\rm kpc}$, halo concentration parameter $c=10$ and spin parameter $\lambda = 0.04$. The stellar bulge follows a~\cite{Hernquist90} profile, with a mass of $3.437 \times 10^9~{\rm M}_\odot$. The stellar disc is of exponential form and has a mass of $4.297 \times 10^{10}~{\rm M}_\odot$, a scale-length of $3.43~{\rm kpc}$, and a scale-height of $0.34~{\rm kpc}$. The gas fraction is $0.18$ and the bulge-to-stellar disc ratio is $0.125$. The star particle mass is $3.437 \times 10^5~{\rm M}_\odot$, the dark matter particle mass is $1.254 \times 10^7~{\rm M}_\odot$, and the median gas cell mass is $859~{\rm M}_\odot$.

We evolve the Milky Way-like initial condition using the moving-mesh hydrodynamics code {\sc Arepo}~\citep{Springel10}. For our median gas cell mass of $859~{\rm M}_\odot$, we set a minimum gravitational softening length of $20$~pc, and employ the adaptive gravitational softening scheme in {\sc Arepo} with a gradation of $1.5$ times the Voronoi gas cell size. We rely on this adaptive softening scheme, along with the fact that our simulation resolves the gas disc scale-height and Toomre mass at all spatial scales, to avoid the majority of artificial fragmentation at scales larger than the Jeans length~\citep{Nelson06}. We set the same softening length of $20$~pc for the stellar particles, and choose a softening length of $260~{\rm pc}$ for the dark matter particles, according to the convergence tests of~\cite{2003MNRAS.338...14P}.

We model the chemical and thermal state of the gas in our simulation according to the chemical network of~\cite{NelsonLanger97,GloverMacLow07a,GloverMacLow07b,Glover10}, which uses a simplified set of reactions to follow the abundances of ${\rm H}$, ${\rm H}_2$, ${\rm H}^+$, ${\rm C}^+$, ${\rm CO}$, ${\rm O}$ and ${\rm e}^-$, while fixing the abundances of helium, silicon, carbon and oxygen to their solar values ($x_{\rm He} = 0.1$, $x_{\rm Si} = 1.5 \times 10^{-5}$, $x_{\rm C}=1.4 \times 10^{-4}$ and $x_{\rm O} = 3.2\times 10^{-4}$, respectively). The initial gas temperature is set to $10^4$~K, and this re-equilibriates on a time-scale of $\lessapprox 10$~Myr to a state of thermal balance between line-emission cooling and heating due to the photoelectric emission from dust grains and polycyclic aromatic hydrocarbons, as they interact with the background interstellar radiation field (ISRF). We set the strength of the ISRF to $1.7$~\cite{Habing68} units as per~\cite{Mathis83}, and the cosmic ray ionisation rate to a value of $2 \times 10^{-16}~{\rm s}^{-1}$~\citep{Indriolo&McCall12}.

\begin{figure*}
	\includegraphics[width=\linewidth]{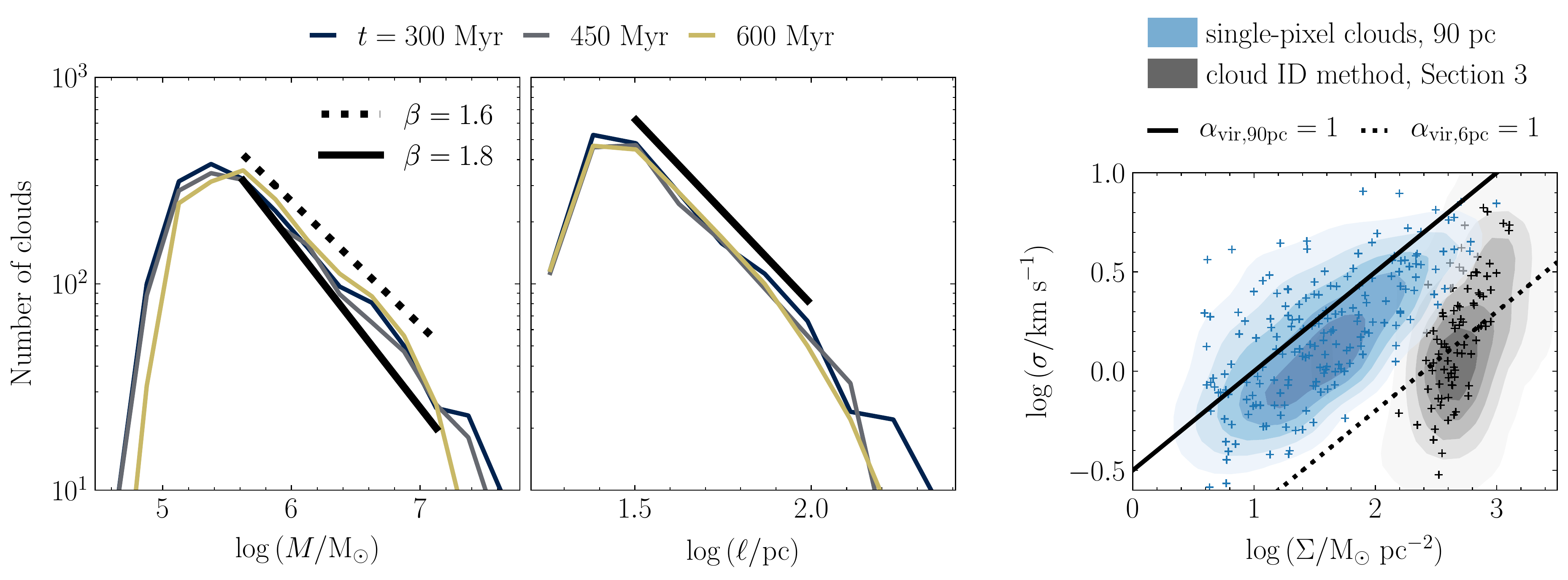}
	\caption{\textit{Left panel:} Mass distribution of the molecular clouds identified at simulation times of $300$, $450$ and $600$~Myr. The solid black and dashed black lines indicate the range of power-law slopes for the observed molecular cloud mass distribution in the Milky Way, given by $\dd N/\dd M \propto M^{-\beta}$, $\beta \in [1.6, 1.8]$. \textit{Centre panel:} Size distribution of the molecular clouds. Here the black line indicates the power-law slope of the observed cloud size distribution in the Milky Way, given by $\dd N/\dd \ell \propto \ell^{-\beta_\ell}$ with $\beta_\ell \sim 2.8$. \textit{Right panel:} Line-of-sight velocity dispersion $\sigma$ as a function of the molecular cloud surface density $\Sigma$, at a simulation time of $600$~Myr. The normalised parameter-space density of the clouds is enclosed by the contours and $1/10$th of the identified clouds are shown as crosses. A virial parameter of $\alpha_{\rm vir}=1$ for spherical beam-filling size $6$~pc (equal to the map resolution) is given by the solid line.}
	\label{Fig::diagnostics}
\end{figure*}

\begin{figure}
	\includegraphics[width=\linewidth]{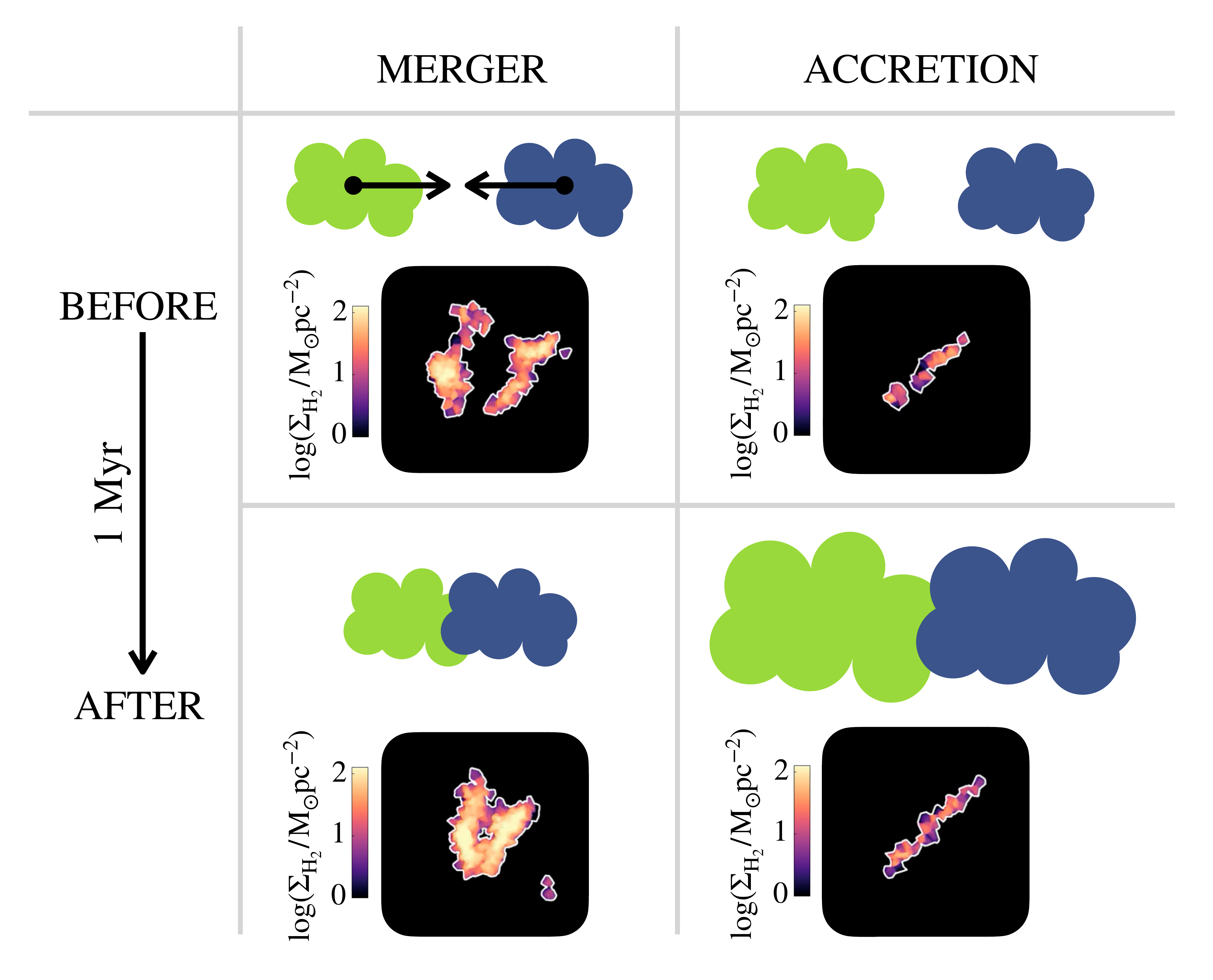}
	\caption{Schematics illustrating the case of two molecular clouds joining in a \textit{merger} (left-hand side) vs. the cases we consider to be \textit{accretion} (right-hand side). Examples of each case, taken from molecular gas surface density maps in our simulation, are shown in the windows below the schematics. If clouds do not approach each other at some relative velocity, but simply grow into each other due to accretion, this is not considered a merger for the purposes of this study.}
	\label{Fig::merger-schematic}
\end{figure}

We form stars in our simulation by locally reproducing the observed relation between the star formation rate and gas surface densities~\citep{Kennicutt98}. The star formation rate density in gas cell $i$ with volume density $\rho_i$ is given by
\begin{equation} \label{Eqn::SF-law}
\frac{\dd \rho_{*,i}}{\dd t} = 
\begin{cases}
      \frac{\epsilon_{\rm ff} \rho_i}{t_{{\rm ff},i}}, \; \rho_i \geq \rho_{\rm thres} \\
      0, \; \rho_i < \rho_{\rm thres}\\
   \end{cases},
\end{equation}
with a local free-fall time-scale $t_{{\rm ff},i} = \sqrt{3\pi/(32 G\rho_i)}$ and a density threshold $\rho_{\rm thresh} = 1000~{\rm cm}^{-3}$, above which star formation is allowed. Our value of $\rho_{\rm thresh}$ corresponds to the gas density at which the Jeans instability sets in at our mass resolution, assuming a maximum gas temperature of $100~{\rm K}$, so that the majority of our star-forming gas is collapsing. The star formation efficiency $\epsilon_{\rm ff}$ is set at 10~per~cent, which is consistent with the upper end of the observed range in dense, molecular gas~\citep{1974ApJ...192L.149Z,Krumholz&Tan07,2012ApJ...745...69K,2014ApJ...782..114E,2016A&A...588A..29H}.

Each star particle formed according to Equation (\ref{Eqn::SF-law}) is assigned a `cluster' of $N$ stars drawn randomly from a~\cite{2005ASSL..327...41C} initial stellar mass function (IMF) using the Stochastically Lighting Up Galaxies model~\citep[SLUG,][]{daSilva12,daSilva14,Krumholz15}. The cluster size $N$ is drawn from a Poisson distribution of expectation value $m_{\rm birth}/\overline{M}$, where $m_{\rm birth}$ is the birth mass of the star particle and $\overline{M}$ is the mean mass of a single star in the cluster. For each cluster, SLUG follows the evolution of individual stars along Padova solar metallicity tracks~\citep{Fagotto94a,Fagotto94b,VazquezLeitherer05} with starburst99-like spectral synthesis~\citep{Leitherer1999}, yielding the ionising luminosity for each star particle, the number $N_{*, {\rm SN}}$ of supernovae it has generated and the mass $\Delta m_*$ it has ejected at each time-step.

The values of $N_{*, {\rm SN}}$ and $\Delta m_*$ for each star particle are used to compute the momentum and thermal energy injected by supernovae at each simulation time-step (if $N_{*, {\rm SN}}>0$). In the case that $N_{*, {\rm SN}}=0$, we assume that all mass loss results from stellar winds. At our mass resolution of $859~{\rm M}_\odot$ per gas cell, we do not resolve the energy conserving, momentum-generating phase of supernova blast-wave expansion, so we must explicitly inject the terminal momentum of the blast-wave, following the work of~\cite{KimmCen14}. We use the unclustered parametrisation of the terminal momentum derived from the high-resolution simulations of~\cite{Gentry17}, given by
\begin{equation} \label{Eqn::Gentry17}
\frac{p_{{\rm t}, k}}{{\rm M}_\odot {\rm kms}^{-1}} = 4.249 \times 10^5 N_{j, {\rm SN}} \Big(\frac{n_k}{{\rm cm}^{-3}}\Big)^{-0.06}.
\end{equation}
In the above, $N_{j, {\rm SN}}$ is the total number of supernovae across all star particles for which gas cell $j$ is the nearest neighbour, and $n_k$ is the volume density of any gas cell $k$ that shares a face with the central cell $j$. The terminal momentum is therefore distributed among all facing gas cells $k$, as described in~\cite{2021MNRAS.505.3470J}. We place an upper limit on the terminal momentum according to the conservation of kinetic energy as the blast-wave moves through the facing gas cells.

In addition to supernova feedback, we include pre-supernova feedback from HII regions, accounting for both the radiation pressure-driven and thermal expansion of the ionised gas surrounding young stellar clusters, according to the model of~\cite{2021MNRAS.505.3470J}. Momentum due to gas and radiation pressure is injected into all of the Voronoi gas cells that share a face with the central `host' cell closest to each stellar cluster. The smallest stellar clusters are made up of individual star particles, and larger clusters are formed from groups of star particles with overlapping ionisation front radii, computed via a Friends-of-Friends grouping prescription. The gas cells inside the Str\"{o}mgren radii of these clusters are self-consistently heated and held above a temperature floor of $7000$~K. We rely on the chemical network to ionise the gas in accordance with the thermal energy injected, and so do not explicitly adjust the chemical state of the heated gas cells within our model.

\subsection{Molecular cloud identification and evolution} \label{Sec::evolution-network}
Molecular clouds are identified within two-dimensional maps of the molecular gas surface density $\Sigma_{\rm H_2}$. The maps are created by post-processing the output of our simulation using the chemistry and radiative transfer model {\sc Despotic}~\citep{Krumholz14}, which provides realistic molecular gas abundances via the escape probability formalism (see Appendix~\ref{App::Sigma-H2-maps} for further details). We compute maps of $\Sigma_{\rm H_2}$ at $1$~Myr intervals and at a spatial resolution of $6$~pc, corresponding to the median radius of Voronoi gas cells above the minimum hydrogen atom number density of $\sim 30~{\rm cm}^{-3}$ that is associated with observed giant molecular clouds. We select contiguous regions of CO-bright molecular hydrogen from the trunk of the dendrogram produced by applying the {\sc Astrodendro} package of~\cite{2008ApJ...679.1338R}, and using a minimum surface density threshold of $\log_{10}{(\Sigma_{\rm H_2}/{\rm M}_\odot{\rm pc}^{-2})} > -3.5$, which captures all of the dense, CO-dominated gas that is shielded from dissociation by the background ISRF.\footnote{This lenient threshold corresponds to the break the distribution of $\Sigma_{\rm H_2}$ produced by our chemical post-processing. At densities above the break are gas cells that contain shielded, CO-dominated gas. Below the break are unshielded gas cells that contain a very low abundance of both ${\rm H}_2$ and CO. We find that increasing the cloud identification threshold to $10~{\rm M}_\odot$ affects the total surface area of identified clouds by $<5$~per~cent.}.

Once the population of molecular clouds has been identified at each simulation time-step, we track the evolution of each cloud as a function of time. We use the sets of Voronoi cell positions and velocities associated with each cloud to project its position forward in time by $1$~Myr. Any clouds at this later time that overlap with the time-projected cloud by more than an area of $36~{\rm pc}^2$ in the galactic plane (one pixel in our surface density maps) are considered to represent the next stage in the cloud's evolution. Each cloud can spawn multiple children (`split nodes') or have multiple parents (`merge nodes'). In the case of multiple parents, we use the relative overlap areas of the parent clouds to approximate the fraction of the resultant cloud mass that each contributes. The resulting cloud evolution network is stored using the {\sc NetworkX} package for Python~\citep{NetworkX}.

Finally, we prune the cloud evolution network by removing clouds that are not well-resolved. We remove clouds that span fewer than nine pixels in area (corresponding to a circular-cloud diameter of $18$~pc) and that have a total CO-luminous mass lower than $1.7\times 10^4~{\rm M}_\odot$ (corresponding to 20 shielded, CO-dominated gas cells). Below these thresholds, there may be fewer than 20 gas cells per cloud, which we deem insufficient to compute its physical properties (e.g. internal gas velocity dispersion). We also remove artefacts associated with regions of faint background CO emission. These have unphysically-low cloud masses and velocity dispersions, so are easily removed with lenient cuts of $<0.2~{\rm M}_\odot$ on the cloud mass and $<0.03~{\rm km~s}^{-1}$ on the cloud velocity dispersion, as described in~\cite{2021MNRAS.505.1678J}.

\subsection{Observational checks of simulated clouds} \label{Sec::obs-checks}
In Figure~\ref{Fig::diagnostics}, we check the key observable diagnostics for our identified cloud population. In the left-hand and central panels, we show the mass and size distributions, respectively, of the molecular cloud samples at simulation times of $300$~Myr, $450$~Myr and $600$~Myr (beginning, middle and end of the simulation time considered). We see that the power-law slope of the cloud mass distribution falls within the spread of values $\dd N/\dd M \propto M^{-\beta}, \beta \in [1.6, 1.8]$ that is observed at high spatial resolution in the Milky Way~\citep{Solomon87,Williams&McKee97,Heyer+09,Roman-Duval+10,Miville-Deschenes17,Colombo+19}. Similarly, the power-law slope of the cloud size distribution agrees well with the observed value of $\dd N/\dd \ell \propto \ell^{-\beta_\ell}, \beta_\ell \sim 2.8$~\citep{Colombo+19}.

\begin{figure*}
	\includegraphics[width=\linewidth]{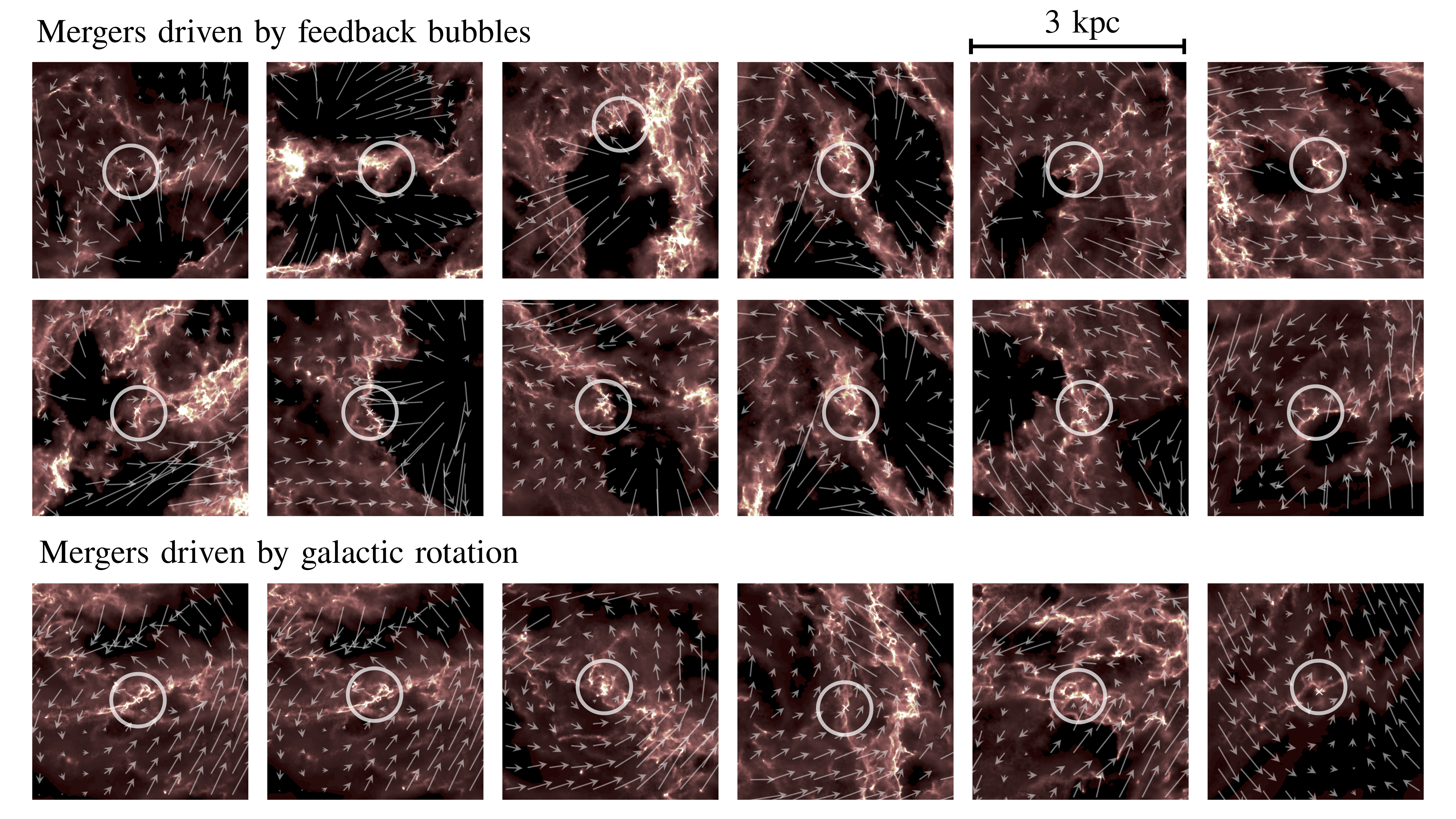}
	\caption{When viewed in the galactic context, the majority of dynamical cloud mergers are positioned at the surfaces of stellar feedback-driven bubbles in the interstellar medium (top 18 panels), or at the centres of solonoidal velocity fields, consistent with the influence of galactic rotation (bottom 6 panels). Some mergers are associated with both driving mechanisms. The panels above show the total gas distribution in the $(3~{\rm kpc})^3$ volume surrounding each merger (white circles) at the instant it occurs. The arrows represent the direction and relative magnitude of the gas velocity field at each position, viewed perpendicular to the galactic mid-plane.}
	\label{Fig::bubble-shear-mergers}
\end{figure*}

In the right-hand panel of Figure~\ref{Fig::diagnostics}, we show the locus of our cloud sample at $600$~Myr (grey contours) in the plane spanned by the cloud surface density $\Sigma$ and the cloud velocity dispersion $\sigma$. The dotted black line represents the virial parameter for spherical beam-filling clouds of size equal to $6$~pc (matching our native map resolution). The clouds in our sample are gravitationally-bound on average (most lie below the dotted line), but a significant fraction are also gravitationally-unbound (above the line).

For the purpose of observational comparison, we also show a sample of molecular clouds (blue contours) identified within the same simulation snapshot, using the technique advocated by~\cite{2016ApJ...831...16L,Sun18}. That is, each pixel with $\log_{10}{(\Sigma_{\rm H_2}/{\rm M}_\odot{\rm pc}^{-2})}>-3.5$ in the map of the molecular hydrogen surface density $\Sigma_{\rm H_2}$ is counted as a separate molecular cloud. For this `single-pixel cloud' sample, we use a map of $\Sigma_{\rm H_2}$ at a spatial resolution of $90$~pc, matching the resolution of the latest observations of molecular gas in nearby galaxies, as presented in~\cite{Sun2020,2021arXiv210407739L}. In accordance with these observational works (see e.g.~Figure 4 of~\citealt{Sun18}), our single-pixel cloud population lies along a line of roughly-constant virial parameter, with a significant fraction of both bound and unbound single-pixel clouds.

By comparing the grey and blue contours, we note that the clouds in our sample (identified via the method outlined in Section~\ref{Sec::evolution-network}) have a significantly steeper slope in the $\Sigma$-$\sigma$ plane than do the clouds in the single-pixel sample. There are two main reasons for this, as follows:
\begin{enumerate}
	\item Via the single-pixel method, the lower-density outskirts of the molecular clouds in our sample are counted as separate clouds. The single-pixel method therefore produces a larger population of low surface-density clouds.
	\item For the molecular clouds in our sample, we apply a lower mass limit of $1.7 \times 10^4~{\rm M}_\odot$. We do not apply this limit to the single-pixel clouds, which allows for clouds with lower masses and surface densities.
\end{enumerate}
We therefore demonstrate that molecular clouds identified as the gas contained within surface density isocontours (as in our sample) may follow different scaling relations to molecular clouds identified via the single-pixel method.

\subsection{Cloud merger sample} \label{Sec::cloudmergersample}
To study the influence of mergers on the evolution of the galactic molecular cloud population (and its star formation), we examine a representative sample of $\sim 600$ `merge-nodes' from our cloud evolution network: points in the network at which two or more clouds are joined to form a single cloud, over a single time-step in the network ($\Delta t = 1$~Myr). The sample of `merge-nodes' are drawn from between galactocentric radii of $2$ and $13$~kpc, and across an interval of $300$~Myr in simulation time. We examine each by eye, both within and perpendicular to the galactic plane, to (1) remove any spurious cases of clouds that do not touch along the galactic $z$-axis (projection effects), and (2) to distinguish mergers from instances of accretion. 

Of these $600$ `merge-nodes', we find that $198$ correspond to dynamically-induced `molecular cloud mergers'. In Figure~\ref{Fig::merger-schematic} we distinguish these mergers (left-hand side) from instances of cloud accretion (right-hand side). We emphasise that `cloud merger' does not refer to clouds that are joined over time by growth and accretion. In this work, we are concerned with elucidating the effect of cloud mergers induced by large-scale dynamical processes such as galactic rotation and shear, or by large (>kpc-scale supernova-driven bubbles. These dynamical mergers must be distinguished by eye, because there is no simple way to quantitatively distinguish a dynamical approach of two clouds from the case that two clouds `grow together' over time. We also note that we require the two merging clouds to have a mass ratio of less than $1/10$. Any joining of two clouds with a more-extreme mass ratio is considered a case of cloud accretion.

\section{Cloud mergers are driven by feedback bubbles and galactic rotation} \label{Sec::driving}
We can now answer the first of our questions concerning cloud mergers in a Milky Way-like galaxy: what physical processes drive cloud mergers, and at what rate?

\subsection{What physical processes drive cloud mergers?} \label{Subsec::driving}
In Figure~\ref{Fig::bubble-shear-mergers}, we illustrate two different scenarios for the driving of cloud mergers in our simulation. The top 18 panels show mergers (with positions indicated by the white target symbols) that occur on the surfaces of feedback-driven bubbles (outlined in white dashed lines, to guide the eye). The grey arrows represent the direction and relative magnitude of the gas velocity field at each position in the galactic mid-plane. We inspect such an image for each merger in our sample of $198$, and find that $\sim 60$~per~cent are positioned at the edges of these bubbles.

The lower six panels in Figure~\ref{Fig::bubble-shear-mergers} illustrate the second large-scale gas distribution that is common to many mergers in our sample. We find that all mergers not associated with feedback bubbles, as well as about $20$~per~cent of the mergers that are associated with feedback bubbles, are positioned near the centres of solonoidal gas velocity fields (indicated by circulating grey arrows in the figure). That is, the mergers occur at the interface of two shearing gas layers. Such mergers can be attributed primarily to large-scale converging gas flows driven by galactic differential rotation.

We therefore conclude that \textbf{the primary drivers of cloud mergers in our simulation are converging gas flows caused by large feedback-driven bubbles, and by galactic differential rotation.}

\subsection{At what rate are cloud mergers driven?}
\label{Sec::merger-rate}

\begin{figure}
	\includegraphics[width=\columnwidth]{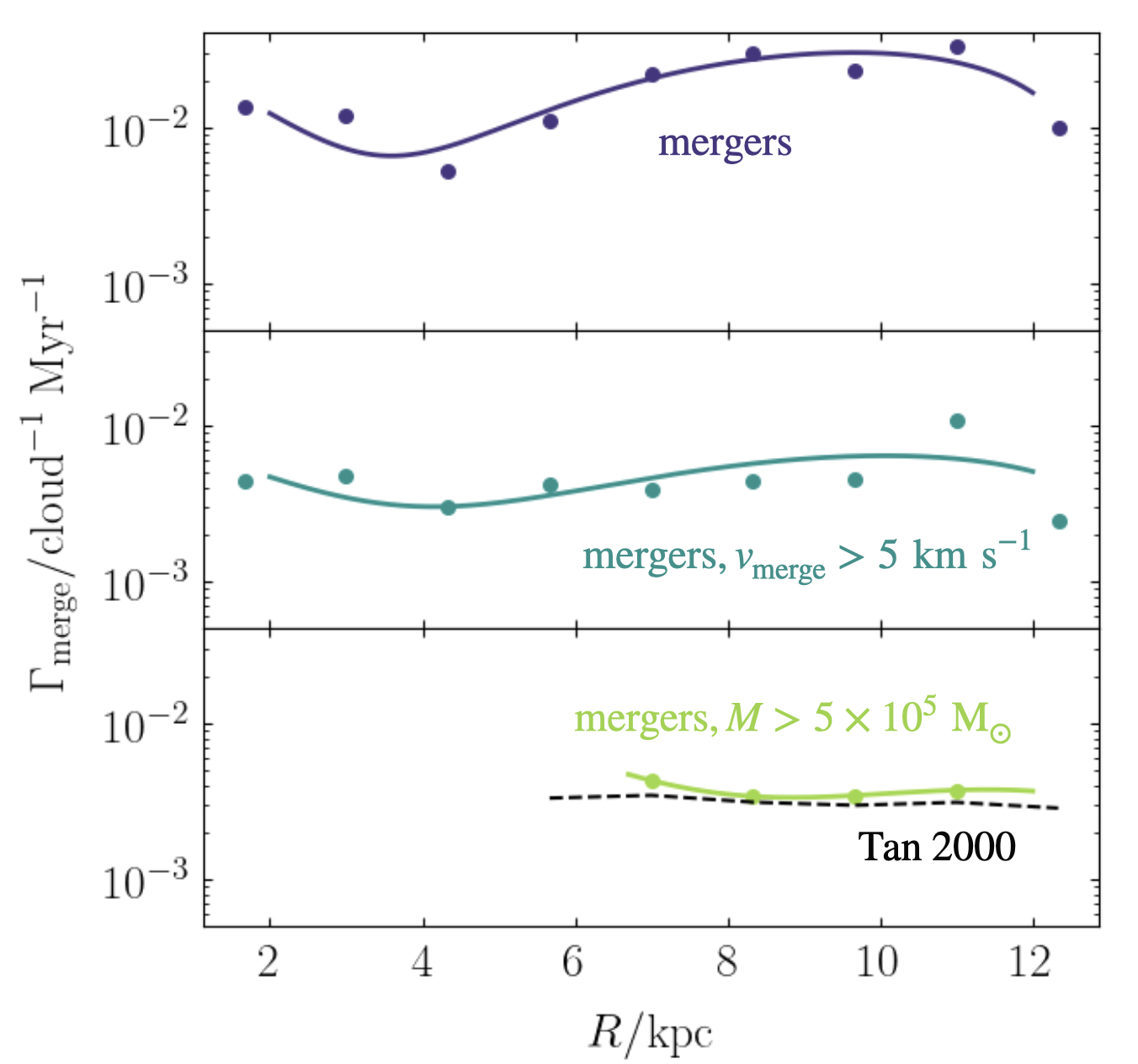}
	\caption{The rate of mergers $\Gamma_{\rm merge}$ per cloud, per million years, as a function of galactocentric radius. The purple line corresponds to the total rate of mergers (top panel), while the turquoise line corresponds to the rate of mergers with relative velocities $v_{\rm merge}>5~{\rm km~s}^{-1}$ (middle panel), which accounts for just 17~per~cent of mergers. The light green line corresponds to mergers of clouds with a minimum mass of $M=5 \times 10^5~{\rm M}_\odot$ (bottom panel), which is the sample of clouds comparable to the merger rate predicted by~\protect\citealt{Tan00} (black dashed line). The merger rate of our massive clouds therefore corresponds well with the analytic prediction for galactic shear-driven collisions (see Section~\protect\ref{Sec::merger-rate}).}
	\label{Fig::mergerrate}
\end{figure}

In Figure~\ref{Fig::mergerrate} we show the rate of cloud mergers $\Gamma_{\rm merge}$ in our simulation as a function of the galactocentric radius $R$. We show this for the full sample of mergers (upper panel, purple points), and for the sub-sample of mergers that occur at a relative speed of $>5~{\rm km~s}^{-1}$ (central panel, turquoise points). To guide the eye, we have fitted spline curves of degree 3 to each merger rate. Across all galactocentric radii, we find that mergers occur at an average rate of $1$~per~cent of clouds per Myr. The rate is lowest at $R=4.3$~kpc, dropping to $\sim 0.5$~per~cent of clouds per Myr. At around $R=11$~kpc, the merger rate attains its maximum value of $3$~per~cent of clouds per Myr. At maximum, $1/6$ of the mergers occur at a speed of $>5~{\rm km~s}^{-1}$, higher than the median velocity dispersion ($\sim 3~{\rm km~s}^{-1}$) inside the clouds. Only this small fraction of mergers may feasibly cause shocks to propagate into the clouds.

For reference, we also compare the merger rate for the high-mass ($M>5\times 10^5~{\rm M}_\odot$) clouds in our simulation (lower panel, light green points) to the analytic prediction of~\cite{Tan00} (black dashed line). At galactocentric radii of $R>6~{\rm kpc}$, the number of such clouds in our simulation is sufficient for their merger rate to be well-defined. The model of~\cite{Tan00} assumes that mergers are driven by the combination of galactic differential rotation (galactic shear) and gas motions along the galactic radial direction. In this scenario, clouds at larger galactocentric radii are moving more slowly than those at smaller radii, such that clouds at smaller radii `catch up to' clouds at larger radii. A non-zero radial gas velocity dispersion may then produce an interaction between the radially-separated clouds. With this reasoning, the merger rate is given by
\begin{equation}
\Gamma_{\rm merge} \sim \frac{2v_s(\sim 1.6 r_t)}{\lambda_{\rm mfp}},
\end{equation}
where $v_s(\sim 1.6 r_t)$ is the relative velocity of clouds separated by a tidal radius $r_t$ set by $M_{\rm min}$, due to galactic differential rotation. The mean free path for a cloud to catch up to another cloud at larger $R$, or to be caught up with by a cloud at smaller $R$, is $\lambda_{\rm mfp}$. Assuming a relatively flat rotation curve, and that the radial velocity dispersion of the clouds results from gravitational torquing as in~\cite{1991ApJ...378..565G}, the author arrives at the merger rate
\begin{equation}
\Gamma_{\rm merge} \sim \frac{9.4 f_{\rm G} \Omega (1+0.3\beta)(1-\beta)}{2\pi Q}.
\end{equation}
The parameter $f_{\rm G}$ represents the `probability of collision' associated with any single encounter between clouds, while $\Omega$ is the galactic orbital angular velocity, $\beta = \dd \ln{v_{\rm c}}/\dd \ln{R}$ is the galactic shear parameter for a circular velocity of $v_{\rm c}(R)$, and $Q$ is the~\cite{Toomre64} gravitational stability parameter. We refer the reader to the cited work for the full analytic derivation.

\begin{figure*}
	\includegraphics[width=\linewidth]{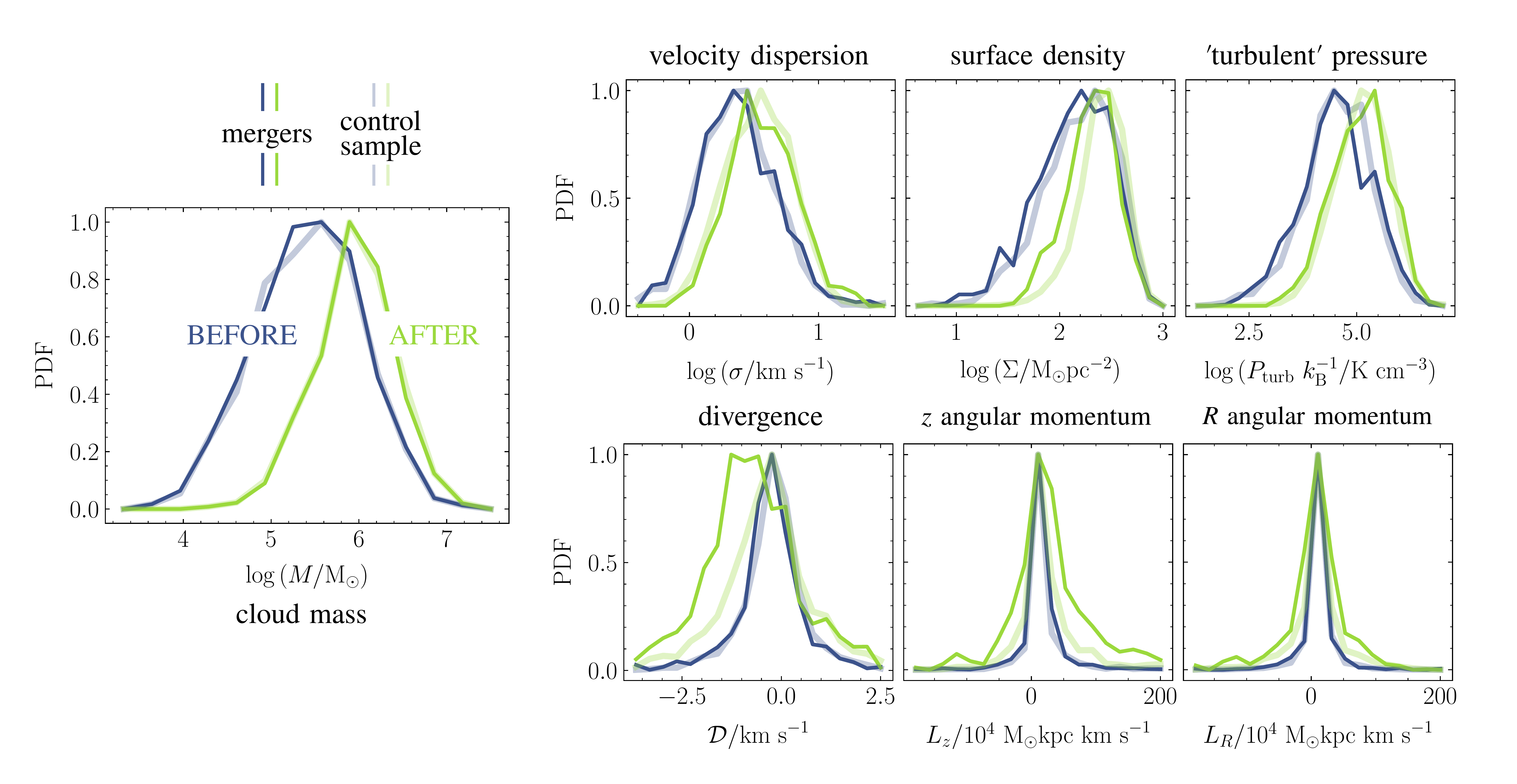}
	\caption{The normalised distributions of physical molecular cloud properties before mergers (in solid blue) and after mergers (in solid green). The transparent lines show a sample of clouds that undergo no mergers throughout their evolution, but which have the same mass distributions as the merging samples (`control sample', see Section~\ref{Sec::before-and-after}).}
	\label{Fig::before-after-mergers}
\end{figure*}

Figure~\ref{Fig::mergerrate} indicates that our high-mass cloud merger rate agrees well with the prediction of~\cite{Tan00}, which depends on the degree of galactic differential rotation and on the radial velocity dispersion between cloud centroids. This agreement implies that on galactic scales, the stirring of molecular gas by supernovae (and the resultant cloud mergers) can be approximately quantified by the radial velocity dispersion of the total galactic gas reservoir, which enters via the Toomre $Q$ parameter in~\cite{Tan00}. Indeed, a typical feedback-blown bubble in our simulation expands to a radius of $\sim 2$~kpc over half an orbital time ($\sim 140$~Myr), giving an approximate expansion velocity of $16~{\rm km~s}^{-1}$. This is comparable to the velocity dispersion of the total gas distribution in our simulation (see Figure 10 of~\citealt{2021MNRAS.505.3470J}).

We conclude that \textbf{mergers occur at an average rate of 1~per~cent of clouds per Myr in our simulation. For a typical cloud lifetime of 20~Myr, this means that one in five clouds would undergo a merger at some point during its lifetime. Mergers are typically slow, occurring at relative speeds of $<5$~km/s. They are unlikely to cause shocks to propagate into the resultant, merged cloud.}

\section{The primary effect of mergers is to aggregate molecular mass} \label{Sec::before-and-after}
The finding in Section~\ref{Sec::merger-rate} that 80~per~cent of mergers in our simulation occur at a relative velocity of $<5~{\rm km~s}^{-1}$ implies that the major impact of mergers on the cloud population is to aggregate mass into larger molecular complexes, rather than to generate shocks that propagate through the resultant merged cloud and alter its properties. The hypothesis that mass aggregation is the major effect of dynamical mergers in flocculent galaxies has been put forward by~\cite{Dobbs15} and~\cite{2021MNRAS.505.1678J}, and will be tested explicitly here.

To separate the effects of mass aggregation from any other merger-induced changes in the properties of merging molecular clouds, we have generated four samples of internal molecular cloud properties using the $198$ identified examples of cloud mergers in our simulation, as follows:
\begin{enumerate}
	\item Cloud properties measured from $3~{\rm Myr}$ before a merger up to $1~{\rm Myr}$ before the merger itself, inclusive (blue solid lines, Figure~\ref{Fig::before-after-mergers}).
	\item Cloud properties measured at the time of the merger, up to $2~{\rm Myr}$ after that merger, inclusive (green solid lines, Figure~\ref{Fig::before-after-mergers}).
	\item `Control sample' of cloud properties sampled at random times from the evolution of clouds that do not undergo any mergers. The sample is selected to have an identical mass distribution to (i) (blue transparent lines, Figure~\ref{Fig::before-after-mergers}).
	\item `Control sample' of cloud properties sampled at random times from the evolution of clouds that do not undergo any mergers. The sample is selected to have an identical mass distribution to (ii) (green transparent lines, Figure~\ref{Fig::before-after-mergers}).
\end{enumerate}
By comparing samples (i) and (ii), we can pin down the changes in internal molecular cloud properties caused by mergers. We have chosen a time interval of $2$~Myr before and after each merger for this comparison, to separate the effect of merger-induced mass aggregation from accretion due to gravitational instability, which occurs for both merging and non-merging clouds. We find that this gravitational accretion, unrelated to cloud mergers, occurs at a rate of $\sim 2 \times 10^4~{\rm M}_\odot~{\rm Myr}^{-1}$ for a simulated, non-merging cloud of median mass $2 \times 10^5~{\rm M}_\odot$, and therefore approaches half of the median merger-induced mass increase over time intervals of $>4~{\rm Myr}$. Our cut-off at $\pm 2~{\rm Myr}$ is the maximum time interval that still ensures that the majority of accretion captured in our analysis is due to cloud mergers. An upper limit of $2~{\rm Myr}$ after each merger also captures any compression-induced increase in the turbulent velocity dispersion of the molecular gas, before it decays on a cloud crossing time~\citep[e.g.][]{1999ApJ...513..259O}. We also exclude periods of cloud evolution before the preceding cloud interaction and after the subsequent cloud interaction, to examine the effects of single mergers.

By comparing sample (i) to the control sample (iii), and (ii) to the control sample (iv), we can determine whether any changes in molecular cloud properties due to mergers are due solely to mass aggregation. In this case, (i) should have the same distribution as (iii) and (ii) should have the same distribution as (iv). Differences between the merger samples and the control samples indicate that the molecular cloud properties are altered independently of merger-induced mass aggregation.

\begin{table*} \label{Tab::before-after-mergers}
\begin{center}
  \caption{Medians and interquartile ranges of cloud properties before and after mergers, along with the mathematical definitions of each cloud property in terms of the properties of the gas cells $i$ in each cloud. Masses $M_i$ denote the molecular masses of gas cells $i$, while angled brackets denote molecular mass-weighted averages over all gas cells in each cloud. The quantity $\ell$ indicates the molecular cloud diameter.}
  \begin{tabular}{@{}l c c c c c@{}}
  \hline
  \textbf{Cloud property} & Definition & Median before & Median after & Intqtl. range before & Intqtl. range after \\
  \hline
  $M/{\rm M}_\odot$ & $\sum_i{M_i}$ & $2.1 \times 10^5$ & $6.0 \times 10^5$ & $[7.1\times 10^4, 5.0\times 10^5]$ & $[2.9\times10^5, 1.2\times 10^6]$ \\
  $\sigma/{\rm km~s}^{-1}$ & $\langle |v_i - \langle v_i \rangle|^2 \rangle$ & $2.0$ & $3.2$ & $[1.3, 3.3]$ & $[2.1, 4.9]$ \\
  $\Sigma/{\rm M}_\odot {\rm pc}^{-2}$ & $M/{\rm Area}$ & $136.3$ & $184.4$ & $[71.5, 235.7]$ & $[122.5, 257.1]$ \\
  $P_{\rm turb} k_{\rm B}^{-1}/{\rm K~cm}^{-3}$ & $61.3 \text{ K cm}^{-3} \left( \frac{\Sigma}{\text{M}_\odot \text{pc}^{-2}}\right) \left( \frac{\sigma}{\text{km s}^{-1}}\right)^2 \left( \frac{\ell}{40 \text{pc}}\right)^{-1}$ & $2.5 \times 10^{4}$ & $8.7 \times 10^{4}$ & $[7.7\times 10^3, 9.2\times10^4]$ & $[3.1\times 10^4, 2.3\times10^5]$ \\
  $\mathcal{D}/{\rm km~s}^{-1}$ & $\langle v_{r, i} \rangle$ & $-0.43$ & $-0.90$ & $[-0.80, -0.02]$ & $[-1.60, -0.08]$ \\
  $L_z / 10^4 {\rm M}_\odot {\rm kpc}~{\rm km~s}^{-1}$ & $\vec{L} = m_i \langle \vec{r}_i \times \vec{v}_i \rangle$ & $0.01$ & $0.43$ & $[-0.03, 0.23]$ & $[-0.31, 4.14]$ \\
  \hline
  \end{tabular}
  \end{center}
 \end{table*}

Given that cloud samples (i) and (ii) correspond to the blue and green solid lines in Figure~\ref{Fig::before-after-mergers}, respectively, we can see that the occurrence of mergers triples the median cloud mass from $2.1$ to $6 \times 10^5~{\rm M}_\odot$ (left-hand side). The median velocity dispersion $\sigma$ and surface density $\Sigma$ are increased by 50~per~cent each, tripling the internal cloud pressure from $P_{\rm turb} = 2.5$ to $8.7 \times 10^4~k_{\rm B}^{-1} {\rm K~cm}^{-3}$ (top row of panels). The definition of each of these physical quantities in terms of the properties of simulated gas cells, along with their median values and interquartile ranges before and after the cloud mergers, are given in Table~\ref{Tab::before-after-mergers}.

Comparing the cloud merger distributions (solid lines) to the control sample distributions (transparent lines) in $\sigma$, $\Sigma$ and $P_{\rm turb}$ demonstrates that these increases in the levels of internal gas turbulence inside the clouds are due primarily to mass aggregation. Higher-mass clouds tend to achieve higher densities in their interiors (due either to compression or collapse) and to attain higher internal gas velocity dispersions in response.

In the lower row of panels in Figure~\ref{Fig::before-after-mergers}, we show the only molecular cloud properties that are altered by mergers, independently of mass aggregation. The first of these (lower left) is the radial divergence $\mathcal{D}$ of gas cell velocities inside the clouds, which is negative if the cloud is contracting towards its centre of mass ($\mathcal{D}<0$) and positive if it is expanding away from its centre of mass ($\mathcal{D}>0$). Clouds that have undergone mergers have a net median compression of $\mathcal{D} = -0.9~{\rm km~s}^{-1}$: double the contraction velocity of clouds that have not undergone mergers. This can be understood simply as the result of the net relative velocity of the merging cloud centroids, which generates a converging flow towards the centre-of-mass of the resultant cloud.

The second cloud property that is altered independently of mass aggregation is the net angular velocity of the gas in the cloud $L_z$, about an axis perpendicular to the galactic mid-plane. In particular, the resultant merged clouds tend to be spun up in the prograde direction $L_z>0$, relative to their unmerged precursors. Like the increased convergence towards the cloud centre of mass, this increased spin likely results from the net angular momentum of the merging cloud system, which is transferred to the merged cloud. The fact that it is more often prograde than retrograde indicates that clouds more often approach each other along the direction of galactic rotation (as expected if the Coriolis effect dominates) rather than antiparallel to it (as would be expected if galactic shear were to dominate).

We therefore conclude that \textbf{mass aggregation accounts for all changes in the internal turbulent properties of molecular clouds due to cloud mergers. Merged clouds have higher masses, therefore higher densities internal turbulent velocity dispersions and pressures.} Independently of mass aggregation, merged clouds have an increased degree of streaming towards their centres of mass, and an increased level of angular momentum, associated with the kinematics of the two-cloud system before the merger occurs.

\section{Cloud mergers aggregate fifty per cent of the gas in the highest-mass clouds} \label{Sec::merger-influence}
\begin{figure*}
	\includegraphics[width=\linewidth]{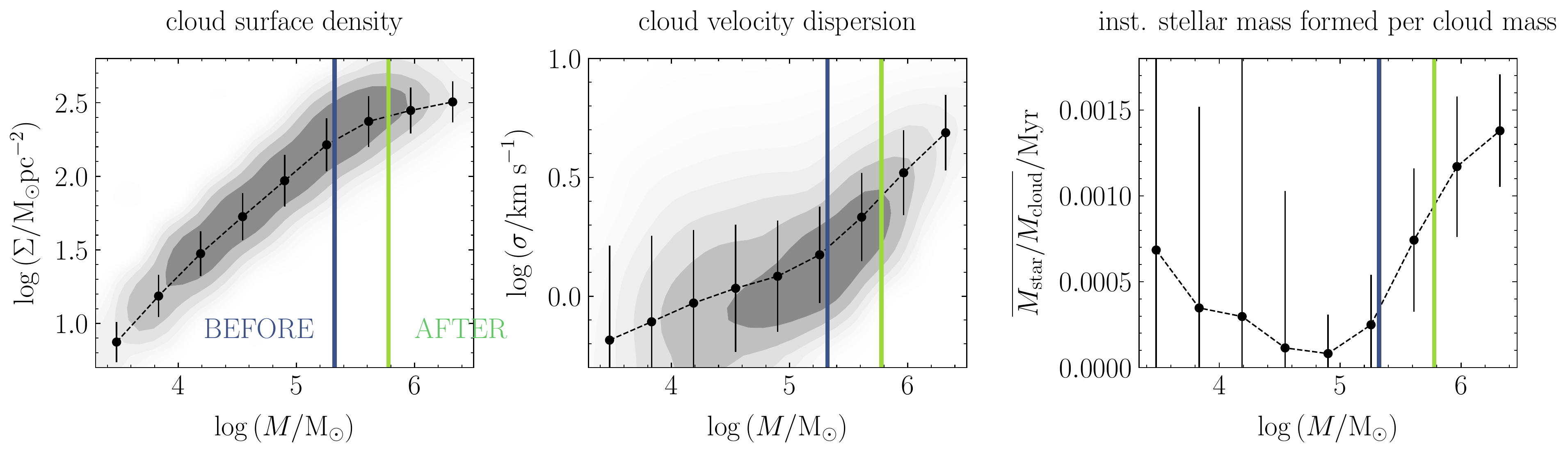}
	\caption{Median molecular cloud surface density (left), median cloud velocity dispersion (centre) and mean instantaneous stellar mass formed per unit cloud mass (right) for the simulated molecular cloud population over the interval of simulation times from $300$ to $600$~Myr. The error-bars represent the standard deviation of the values for each mass bin, and we have added a black dashed line to each dataset to guide the eye. The blue and green vertical lines denote the mean molecular cloud masses of merging clouds before and after their mergers, respectively.}
	\label{Fig::cloud-props-vs-mass}
\end{figure*}

\begin{figure*}
	\includegraphics[width=\linewidth]{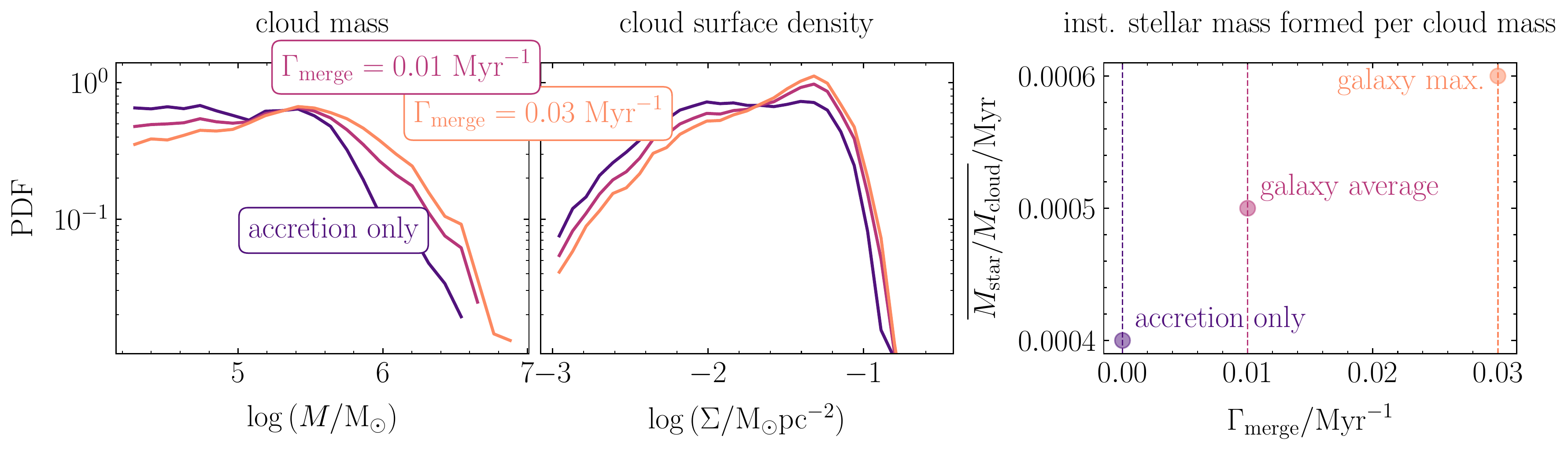}
	\caption{Molecular cloud mass distribution (left), surface density distribution (center) and mean instantaneous stellar mass produced per unit cloud mass per Myr (right) for the entire simulated cloud sample (magenta lines/point), the simulated cloud sample with all clouds removed that have undergone mergers (purple lines/point), and the simulated cloud sample with some non-interacting clouds removed, to mimic a cloud merger rate of $\Gamma_{\rm merge}=0.03~{\rm Myr}^{-1}$ (orange lines/point).}
	\label{Fig::galaxy-SFR-props}
\end{figure*}

Having found that the primary effect of cloud mergers is to aggregate molecular gas into higher-mass clouds, we can now answer our final question from Section~\ref{Sec::Introduction}: what would happen to the galactic distributions of molecular cloud masses, densities and star formation efficiencies \textit{without} the presence of mergers?

In Figure~\ref{Fig::cloud-props-vs-mass}, we show the median values of the cloud surface density (left-hand panel), velocity dispersion (central panel) and instantaneous star formation efficiency (right-hand panel) over an interval of $1$~Myr, as a function of the cloud mass, for all molecular clouds in our simulated galaxy. The vertical blue and green lines indicate the mean mass of merging clouds before and after the merger has occurred, as in Figure~\ref{Fig::before-after-mergers}. We see that higher-mass clouds have higher surface densities and velocity dispersions than do lower-mass clouds---a generalisation of the result shown in Figure~\ref{Fig::before-after-mergers} for the merging clouds. This increase in density with mass corresponds to a steep increase in the instantaneous star formation efficiency (by a factor of $\sim 15$) as the cloud mass is increased from $10^5~{\rm M}_\odot$ up to $2 \times 10^6~{\rm M}_\odot$. That is, the increased self-gravitation inside higher-mass clouds leads to an increased internal pressure, so that higher-mass clouds convert gas into stars at a significantly-faster rate than do low-mass clouds.

In Figure~\ref{Fig::galaxy-SFR-props}, we demonstrate how the galactic rate of molecular cloud mergers alters the cloud mass distribution (left-hand panel), with follow-on consequences for the distribution of surface densities (central panel) and the galactic star formation efficiency (right-hand panel). The three line/data-point colours represent the following three samples of clouds:
\begin{enumerate}
	\item Magenta, galaxy average: The entire population of clouds in the simulated galaxy, with a merger rate of $\Gamma_{\rm merge} = 0.01~{\rm cloud}^{-1}~{\rm Myr}^{-1}$ (see Figure~\ref{Fig::mergerrate}).
	\item Purple, accretion only: A modified galactic cloud population in which we have replaced all periods of cloud evolution after mergers with time-directed segments of cloud evolution that begin with the same masses as the unmerged clouds, but are sampled only from the non-interacting cloud population. This sample mimics the cloud population of a galaxy in which no mergers occur ($\Gamma_{\rm merge}=0~{\rm cloud}^{-1}~{\rm Myr}^{-1}$).
	\item Orange, galaxy maximum: A modified galactic cloud population in which we have removed a number of the non-interacting clouds to mimic a merger rate of $\Gamma_{\rm merge}=0.03~{\rm cloud}^{-1}~{\rm Myr}^{-1}$. This corresponds to the maximum merger rate achieved in our simulation, at any galactocentric radius ($\Gamma_{\rm merge}=0.03~{\rm cloud}^{-1}~{\rm Myr}^{-1}$, see Figure~\ref{Fig::mergerrate}).
\end{enumerate}
Comparing the cloud mass distributions for the cloud samples with galaxy-average (magenta line, Figure~\ref{Fig::galaxy-SFR-props}) and accretion-only (purple line) merger rates, we find that the occurrence of cloud mergers in the simulation accounts for over 50~per~cent of the molecular gas mass aggregated into the highest-mass clouds (clouds with masses $M>2 \times 10^6~{\rm M}_\odot$). In the cloud sample with accretion only, 11~per~cent of the galactic molecular gas reservoir is in these high-mass clouds: this increases to 23~per~cent at a merger rate of $0.01~{\rm cloud}^{-1}~{\rm Myr}^{-1}$, and to 30~per~cent at the maximum merger rate.

Clouds with masses above $2 \times 10^6~{\rm M}_\odot$ account for 50~per~cent of the star formation in our simulation, and so this merger-induced increase in the fraction of high-mass clouds corresponds to a 25~per~cent increase in the instantaneous star formation efficiency, from $4 \times 10^{-3}$ up to $5 \times 10^{-3}$ over $1$~Myr (magenta data-point, right-hand panel), relative to the accretion-only case (purple data-point, right-hand panel). At the maximum cloud merger rate of $\Gamma_{\rm merge}=0.03~{\rm cloud}^{-1}~{\rm Myr}^{-1}$, the efficiency is increased by 50~per~cent relative to the case of no mergers, up to $6 \times 10^{-3}$ over $1$~Myr.

From this analysis we can conclude that \textbf{cloud mergers account for 50~per~cent of the molecular mass contained in the highest-mass clouds in our simulation. They therefore raise the star formation efficiency in the galactic molecular gas reservoir by 25~per~cent, relative to the case in which the cloud population is produced by accretion alone.}

\section{Discussion} \label{Sec::discussion}
\subsection{Comparison to previous work}
The results of our analysis are in broad agreement with the analytic theory of~\cite{2015A&A...580A..49I} and~\cite{2017ApJ...836..175K}, which describes the evolution of the number density of molecular clouds at a given mass according to a cloud accretion, dispersal, and coagulation due to mergers. In particular, the influence of cloud mergers is examined using an accretion time-scale of $10$~Myr and a dispersal time-scale of $14$~Myr. Although we do not explicitly examine these time-scales in the present work, we note that the simulation we present here is similar in morphology to the simulated galaxies presented in~\cite{2021MNRAS.505.1678J}. Both of the fiducial time-scales used in~\cite{2017ApJ...836..175K} are an adequate approximation of the mean time-scales for accretion and dispersal measured in that work. We can therefore roughly compare our maximum molecular cloud masses with and without mergers (magenta and purple lines, respectively, in Figure~\ref{Fig::galaxy-SFR-props}) to the maximum masses that can be inferred from the black lines (converged distributions of the cloud number density with mass) in Figures 2 and 1, respectively, of~\cite{2017ApJ...836..175K}.

Quantitatively, we see that the maximum cloud mass in our simulation is increased from $\sim 10^{6.5}$ to $\sim 10^{6.75}~{\rm M}_\odot$ by the inclusion of molecular cloud mergers, very close to the increase reported in Figures 2 and 1 of~\cite{2017ApJ...836..175K}. Qualitatively, we can also see that the major difference in the slope of the cloud mass function due to the mergers in our simulation is at high cloud masses $M \ga 10^{5.5}~{\rm M}_\odot$, in agreement with the results of~\cite{2017ApJ...836..175K}.

Our numerical study of cloud mergers within an isolated galaxy simulation is also closely-comparable to the work presented in~\cite{2015MNRAS.446.3608D}. In that work, the overall rate of cloud mergers for a flocculent Milky Way-like galaxy is found to be $0.04~{\rm cloud}^{-1}~{\rm Myr}^{-1}$, or once in $28~{\rm Myr}$. This is quadruple the rate of dynamical mergers in our sample ($0.01~{\rm cloud}^{-1}~{\rm Myr}^{-1}$) but close to the frequency of merge-nodes (dynamical or otherwise) in our cloud evolution network. It is therefore possible that the sample of mergers found by~\cite{2015MNRAS.446.3608D} also contains a significant fraction of non-dynamical mergers (clouds grow together, but their centres of mass do not approach each other). The mergers shown in their Figures 12 and 14 appear to be such cases.

Although~\cite{2015MNRAS.446.3608D} do not study the statistical properties of cloud mergers, they find that the mergers presented in their Figures 12 and 14 are typical of the mergers in their simulation. That is, there are few dynamical mergers of the kind presented in our Figure~\ref{Fig::merger-schematic}, and mergers appear not to occur at the edges of feedback-driven bubbles, as presented in our Figure~\ref{Fig::bubble-shear-mergers}.

The higher rate of dynamical (and in particular feedback-driven) mergers in our simulation may be explained by the fact that our stellar feedback is significantly more violent than the feedback in the galaxies of~\cite{2015MNRAS.446.3608D}. It blows much larger holes in the interstellar medium and produces significant outflows, as well as promoting the onset of flocculent spiral arms (see Figure~\ref{Fig::morphology}). This difference in the morphology of our interstellar medium relative to~\cite{2015MNRAS.446.3608D} may be explained by the longer delay between star formation and supernova explosions in our simulation. The energy and momentum deposited due to the supernovae in~\cite{2015MNRAS.446.3608D} is instantaneous, whereas the supernova explosions in our simulation occur at an interval after star formation that is calculated stochastically according to a~\cite{Chabrier03} initial stellar mass function, and may be $>10$~Myr in some cases~\citep[see][]{2021MNRAS.505.3470J}. Longer delays allow for a higher degree of star formation and supernova clustering, enhancing the power of the explosions. We therefore expect a significantly higher number of collisions due to expanding feedback bubbles in our simulation.

\subsection{Caveats}
A possible caveat to the analysis presented in Section~\ref{Sec::merger-influence} arises due to the star formation prescription used in our simulation, outlined in Section~\ref{Sec::sims}. Stars form stochastically in our simulated galaxy, at an efficiency of 10~per~cent above a hydrogen gas density threshold of $1000~{\rm cm}^{-3}$. This means that if gas is compressed to high densities over a short interval of time (i.e.~by a passing shock), our simulation will not recover the increased star formation efficiency expected in the shocked gas \citep[see e.g.][]{2017ApJ...841...88W,2020MNRAS.499.1099L,2020MNRAS.494..246T}. This means that the enhancement of the star formation efficiency in the sample of clouds containing mergers may feasibly be higher than the 25~per~cent presented in Section~\ref{Sec::merger-influence}.

However, given the low relative speed of cloud mergers in our simulation (only 17~per~cent of mergers occurring at speeds $>5~{\rm km~s}^{-1}$, and none at speeds $>10~{\rm km~s}^{-1}$), we do not expect this caveat to significantly affect our results. We have shown in Section~\ref{Sec::merger-rate} that the average speed of mergers in our simulation is comparable to the internal velocity dispersion of the merging clouds, and should therefore be incapable of driving shocks into these clouds. Furthermore, in Section~\ref{Sec::before-and-after} we have shown that the internal motions of the merged clouds in our sample are altered only by the mass aggregated during the merger, indicating that no shock propagates through the resulting cloud. Our star formation prescription is adequate to handle such slow mergers.

\subsection{Future work}
The mergers in our simulation are infrequent (1~per~cent of clouds per Myr) relative to mergers in barred galaxies~\citep{2014MNRAS.445L..65F}, in grand-design spirals~\citep{2015MNRAS.446.3608D} and in galaxy mergers~\citep{2022MNRAS.tmp.1143L}. Despite this, they produce a clear shift in the distribution of cloud masses and of the star formation efficiency of the molecular gas in the galaxy, as shown in our Section~\ref{Sec::merger-influence}. By applying the analysis presented here to simulations of galaxies with higher merger rates, it will be possible to quantify the role of cloud mergers in producing the widely-varying observed properties of molecular gas (and star formation) in these galaxies. In particular, it would be fruitful to apply our analysis to a barred or grand-design spiral galaxy, to determine whether the double-peaked distribution of molecular gas surface densities and velocity dispersions~\citep{Sun18} can be explained by the presence of merging clouds. Our analysis could also be applied to a galaxy merger simulation~\citep[e.g.][]{2022MNRAS.tmp.1143L} to study the impact of very fast cloud mergers on its cloud population.

\section{Conclusions} \label{Sec::conclusions}
We have investigated the physical drivers of molecular cloud mergers in a Milky Way-like isolated galaxy simulation, and determined their role in shaping the properties of the galactic molecular cloud population, as well as the star formation rate in molecular gas. Our main conclusions are as follows:

\begin{enumerate}
	\item Cloud mergers in our flocculent spiral galaxy are slow (83~per~cent have collision speeds below 5~km/s) and are associated primarily with large supernova-driven bubbles blown in the dense gas of the interstellar medium, and with galactic rotation.
	\item The major effect of these slow mergers is to aggregate mass into higher-mass molecular clouds. Merged clouds have significantly higher internal densities, velocity dispersions and instantaneous star formation efficiencies than their unmerged precursors, and these increases can all be reproduced in non-merging clouds of equal mass to the merged clouds.
	\item 50~per~cent of the mass contained in high-mass clouds (above $2 \times 10^6~{\rm M}_\odot$, accounting themselves for 50~per~cent of star formation in the simulated galaxy), is aggregated by mergers, rather than by simple accretion.
	\item Due to this increase in the fraction of molecular mass contained in high-mass clouds, the instantaneous star formation efficiency in our simulation is elevated by 25~per~cent, relative to the star formation efficiency that would be expected for a similar galaxy in which no mergers occur.
\end{enumerate}
We emphasise that although an increase of 25~per~cent in star formation efficiency might seem modest, our simulated flocculent galaxy has a cloud merger rate of just 1~per~cent of clouds per Myr. The bars of grand design spiral galaxies are found to have merger rates as high as 40~per~cent of clouds per Myr. The analysis presented here implies that in such galaxies, dynamically-driven molecular cloud mergers may play a key role in building the high-mass end of the molecular cloud mass function, and therefore in enhancing the galactic star formation efficiency.

\section*{Acknowledgements}
We thank an anonymous referee for a constructive report, which improved the clarity of the text. We thank Volker Springel for providing us access to Arepo. MS would like to acknowledge support from the Harvard College Research Program. SMRJ is supported by Harvard University through the ITC. AG acknowledges NSF Grant 1908419 for supporting this work. This work was undertaken with the assistance of resources and services from the National Computational Infrastructure (NCI; award jh2), which is supported by the Australian Government. We are grateful to Angus Beane, Michael Foley, Vadim Semenov, Enrique V\'{a}zquez-Semadeni, Andrew Winter and Catherine Zucker for helpful discussions.

\section*{Data Availability Statement}
The data underlying this article are available in the article and in its online supplementary material.



\bibliographystyle{mnras}
\bibliography{bibliography} 



\appendix
\section{Calculation of the molecular gas surface density $\Sigma_{\rm H_2}$} \label{App::Sigma-H2-maps}
As noted in Section~\ref{Sec::evolution-network}, we identify molecular clouds in two-dimensional maps of the molecular gas surface density $\Sigma_{\rm H_2}$. To calculate the molecular gas surface density, we post-process the simulation output using the {\sc Despotic} model for astrochemistry and radiative transfer~\citep{Krumholz13a}. At the mass resolution of our simulation, the self-shielding of molecular hydrogen from the ambient UV radiation field cannot be accurately computed during run-time, so that the molecular hydrogen abundance is under-estimated by a factor $\sim 2$, requiring this value to be re-calculated in post-processing. Within {\sc Despotic}, the escape probability formalism is applied to compute the CO line emission from each gas cell according to its hydrogen atom number density $n_{\rm H}$, column density $N_{\rm H}$ and virial parameter $\alpha_{\rm vir}$, assuming that the cells are approximately spherical. In practice, the line luminosity varies smoothly with the variables $n_{\rm H}$, $N_{\rm H}$, and $\alpha_{\rm vir}$. We therefore interpolate over a grid of pre-calculated models at regularly-spaced logarithmic intervals in these variables to reduce computational cost. The hydrogen surface density is estimated via the local approximation of~\cite{Safranek-Shrader+17} as $N_{\rm H}=\lambda_{\rm J} n_{\rm H}$, where $\lambda_{\rm J}=(\pi c_s^2/G\rho)^{1/2}$ is the Jeans length, with an upper limit of $T=40~{\rm K}$ on the gas cell temperature. The virial parameter is calculated from the turbulent velocity dispersion of each gas cell according to~\cite{MacLaren1988,BertoldiMcKee1992}. The line emission is self-consistently coupled to the chemical and thermal evolution of the gas, including carbon and oxygen chemistry~\citep{Gong17}, gas heating by cosmic rays and the grain photo-electric effect, line cooling due to ${\rm C}^+$, ${\rm C}$, ${\rm O}$ and ${\rm CO}$ and thermal exchange between dust and gas. We match the ISRF strength and cosmic ionisation rate to the values used in our live chemistry.

Having calculated values of the CO line luminosity for each simulated gas cell, we compute the CO-bright molecular hydrogen surface density as
\begin{equation}
\begin{split}
\Sigma_{\rm H_2}[{\rm M}_\odot{\rm pc}^{-2}] = &\frac{2.3 \times 10^{-29}{\rm M}_\odot({\rm erg~s}^{-1})^{-1}}{m_{\rm H}[{\rm M}_\odot]} \\
&\times \int^{\infty}_{-\infty}{\dd z^\prime \rho_{\rm g}(z^\prime) L_{\rm CO}[{\rm erg~s}^{-1}~{\rm H~atom}^{-1}]},
\end{split}
\end{equation}
where $\rho_{\rm g}(z)$ is the total gas volume density in ${\rm M}_\odot~{\rm pc}^{-3}$ at a distance $z$ (in ${\rm pc}$) from the galactic mid-plane. The factor of $2.3 \times 10^{-29}~{\rm M}_\odot~({\rm erg~s}^{-1})^{-1}$ combines the mass-to-luminosity conversion factor $\alpha_{\rm CO}=4.3~{\rm M}_\odot~({\rm K}~{\rm kms}^{-1}~{\rm pc}^2)^{-1}$ of~\cite{Bolatto13} with the line-luminosity conversion factor $5.31 \times 10^{-30}({\rm K~kms}^{-1}{\rm pc}^2)/({\rm erg~s}^{-1})$ for the CO $J=1\rightarrow 0$ transition at redshift $z=0$~\citep{SolomonVandenBout05}.


\bsp	
\label{lastpage}
\end{document}